\documentstyle[12pt]{article}
\font\tenrm=cmr10

\begin{document}
% new macro for bibliography
%\renewenvironment{thebibliography}[1]
%  { \begin{list}{\arabic{enumi}.}
%    {\usecounter{enumi} \setlength{\parsep}{0pt}
%     \setlength{\itemsep}{3pt} \settowidth{\labelwidth}{#1.}
%     \sloppy
%    }}{\end{list}}

\parindent=1.5pc

\begin{center}{{\bf
               SPONTANEOUS BREAKING OF CHIRAL SYMMETRY\\
               \vglue 3pt
               AS A CONSEQUENCE OF CONFINEMENT
               }\\
\vglue 1.0cm
{A.~N.~IVANOV, N.~I.~TROITSKAYA}\\
\vglue 0.2cm
\baselineskip=14pt
{\it State Technical University}\\
\baselineskip=14pt
{\it 195251 St. Petersburg, Russia}\\
\vglue 0.4cm
{M.~FABER, M.~SCHALER}\\
\vglue 0.2cm
\baselineskip=14pt
{\it Institut f\"ur Kernphysik, Technische Universit\"at Wien}\\
\baselineskip=14pt
{\it A-1040 Vienna, Austria}\\
\vglue 0.4cm
{M.~NAGY}\\
\vglue 0.2cm
\baselineskip=14pt
{\it Institute of Physics, Slovak Academy of Sciences}\\
\baselineskip=14pt
{\it 842 28 Bratislava, Slovakia}\\
\vglue 2.8cm
{ABSTRACT}}
\end{center}
\vglue 0.8cm
{\rightskip=3pc
 \leftskip=3pc
 \tenrm\baselineskip=18pt
 \noindent
We show that at the leading order in the large-$N$ expansion a lattice QCD
motivated linear rising confinement
potential at large distances
leads to a non-local four-quark interaction that realizes spontaneous
breaking
of chiral symmetry (SBCS) in the same way the Nambu-Jona-Lasinio model
does. The dynamical quark mass $m$, which represents the solution of the
gap-equation, is proportional to the square root of the string tension
$\sigma$ and takes the form at the leading order in the large-$N$ expansion
$m=2 \sqrt{\sigma}/\pi$ with $\sigma=0.27$ GeV$^2$. The Nambu-Jona-Lasinio
phenomenological constant $G_1$, which is responsible for
SBCS, is
expressed in terms of the string tension and the confinement radius.
\vglue 0.8cm}
\vfill
\begin{center}
IK--TUW--Preprint 9305401\\
May 1993
\end{center}
\newpage
\baselineskip=22pt
\parskip=33pt
It is well known that spontaneous breaking of chiral symmetry (SBCS) and
confinement
play an important role for the low-energy interactions of hadrons. In ref.
[1]
a qualitative analysis
of the influence of confinement on SBCS was performed.

In this paper we propose a quantitative description of the connection
between confinement and SBCS. We identify the observed mechanism of SBCS
within QCD with that proposed by Nambu and Jona-Lasinio [2], which stems
from the pheno\-meno\-lo\-gi\-cal local four-quark interaction.
First of all, let us remind the main features of the NJL-mechanism of SBCS.

The total Lagrangian of quarks interacting in the NJL-way takes the form
\begin{equation}
{\cal L}(x)= \bar \psi (x) i \gamma^{\mu} \partial_{\mu} \psi (x) + {\cal
L}_{NJL}(x)
\end{equation}
where ${\cal L}_{NJL}(x)$ is the NJL local four-quark interaction, which is
invariant under
$U(3) \otimes
U(3)$ chiral rotations
[3,4]
\begin{eqnarray}
\label{eins}
{\cal L}_{NJL}(x) &=& 2 G_1 \left[ (\bar \psi (x) t^a \psi (x))^2 +
(\bar \psi (x) \gamma^5 t^a \psi (x))^2 \right]- \nonumber \\
&& -2 G_2 \left[ (\bar \psi (x) \gamma_{\mu} t^a \psi (x))^2 +
(\bar \psi (x) \gamma_{\mu} \gamma^5 t^a \psi (x))^2 \right].
\end{eqnarray}
Here $G_1$ and $G_2$ are positive phenomenological constants with its
numerical values
$G_1 \simeq 10 \mbox{GeV}^{-2}$
and $G_2 \simeq 15 \mbox{GeV}^{-2}$, which lead to the approximate ratio
$G_2/G_1 \simeq 3/2$. Furthermore, $\psi=(\psi_u,\psi_d,\psi_s)$ are
massless
current quark fields with $N$ color degrees of freedom, and $t^a$
($a=1,\dots,8$) denote the
Gell-Mann matrices of the $SU(3)_{flavor}$ group, which are normalized by
the condition
$\mbox{tr}(t^a t^b)= \delta^{ab}/2$. The strong attraction in the $\bar
\psi \psi$-channels
caused by the local four-quark interaction (\ref{eins}) ensures SBCS
and the emergence of four nonets of
$\bar \psi \psi$-collective excitations with the quantum numbers of
low-lying scalar, pseudoscalar,
vector and axial-vector mesons [2-4].

In the one quark-loop approximation the NJL-interaction (\ref{eins})
generates the following contribution to the dynamical quark mass [2,3]
\begin{equation}
\label{zwei}
\delta {\cal L}_{eff}^{(mass)} (x) = -2 G_1 \mbox{tr}_{C,L} [iS_F(0)] \bar
\psi (x) \psi (x)
\end{equation}
where $S_F(x)$ is the Green's function of a Dirac particle with mass $m$,
which is
con\-si\-dered as the
dynamical quark mass. One obtains
\begin{equation}
\label{drei}
\mbox{tr}_{C,L} [iS_F(0)] = \frac{N}{16 \pi^2} \int \frac{d^4
k}{\pi^2 i} \mbox{tr}_L
\left( \frac{1}{m-\gamma^{\mu} k_{\mu}} \right) = 4m I_1(m)
\end{equation}
where $I_1(m)$ is a quadratically divergent integral
\begin{equation}
\label{vier}
I_1(m) =\frac{N}{16 \pi^2} \int \frac{d^4 k}{\pi^2 i} \frac{1}{m^2 -
k^2}=
\frac{N}{16 \pi^2} \left[ \Lambda_{\chi}^2 -m^2 \ln \left(
1+\frac{\Lambda_{\chi}^2}{m^2} \right) \right].
\end{equation}
The cut-off $\Lambda_{\chi}$ plays the role of the scale of SBCS.

Following the NJL-prescription and assuming that the dynamical mass $m$ is
completely determined by the
one quark-loop contribution [2,3] one gets the
gap-equation
\begin{equation}
\label{funf}
m=m\, 8G_1 \,  I_1(m).
\end{equation}
In the case $G_1=0$ the gap-equation has only the trivial solution $m=0$.
For
$G_1\not=0$ there may exist, in addition to the trivial solution $m=0$,
which corresponds to the chirally invariant phase of the quark system with
the local four-quark interaction, the non-trivial solution
$m=\Lambda_{\chi} \varphi
(G_1 \Lambda_{\chi}^2)$. A solution $m\not= 0$
of the gap-equation describes the chirally broken phase of the quark system
with the interaction (\ref{eins}).
In this phase the system is unstable against the creation of $\bar \psi
\psi$-collective excitations with the quantum numbers of low-lying mesons,
and it possesses a ground state, which is the non-perturbative
many-particle vacuum of superconductivity [2].
We gave only a cursory outline
of the main points of the NJL-model. For a detailled acquaintance of the
NJL-mechanism of SBCS and its connection
to BCS-theory of superconductivity we refer the readers to the papers of
Nambu and
Jona-Lasinio, Eguchi and
Kikkawa [2,3].

The brief outline of the NJL-model shows that
in order to
analyse the influence of confinement on SBCS and to derive the
NJL-interaction within QCD we have to regard the four-quark interaction,
which
describes $\bar \psi \psi$- (or $\psi \psi$-) scattering. The general
expression for the action which accounts for
$\bar \psi \psi$- (or $\psi \psi$-) scattering
within QCD can be written in the
following form [5]
\begin{equation}
\label{sechs}
-i S_{QCD}^{(\bar \psi \psi)} =\frac{1}{2} \int d^4 x_1 d^4 x_2
\bar \psi (x_1) \gamma^{\mu} t^A_c \psi (x_1) K_{\mu\nu}^{AB}(x_1,x_2)
\bar \psi (x_2) \gamma^{\nu} t^B_c \psi (x_2) + \dots.
\end{equation}
Here $t^A_c$ ($A=1,\dots,N^2-1$) are the matrices of the $SU(N)_C$ group
normalized by the condition
$\mbox{tr}_C(t^A_c t^B_c)= \delta^{AB}/2$, and $K^{AB}(x_1,x_2)$
denotes a structure function.
For large distances the structure function $K^{AB} (x_1,x_2)$ can be
expressed in terms of
a confinement potential
$V(\vec x_1 - \vec x_2)$ [1]. The dots in (\ref{sechs}) describe further
possible
contributions, which are less important in the confinement regime [1].
Let us now determine the r.h.s. of
eq. (\ref{sechs}) at large distances [1]
\begin{eqnarray}
\label{sieben}
-i S_{QCD}^{(\bar \psi \psi)} &=& -\frac{1}{2} \int d^3 x_1 d^3 x_2
\int_{-\infty}^{\infty} dt \,
\bar \psi (\vec x_1,t) \gamma^{0} t^A_c \psi (\vec x_1,t) \times \nonumber
\\
&& \times V(\vec x_1 - \vec x_2)
\bar \psi (\vec x_2,t) \gamma^{0} t^A_c \psi (\vec x_2,t) .
\end{eqnarray}
In the following, we consider a linear rising confining potential $V(\vec
x_1 - \vec
x_2)$ of the form
\begin{equation}
\label{acht}
V(\vec x_1 - \vec x_2) = \sigma \, | \vec x_1 - \vec x_2 | -2 \sqrt{\sigma}
\end{equation}
where $\sigma $ is the string tension with the  numerical
value $\sigma \simeq 0.27$ $\mbox{GeV}^2$ [6]. The potential (\ref{acht})
is in excellent agreement with the experimental results of meson
spectroscopy [7]. It should be noted that this potential can be calculated
numerically and in the strong-coupling expansion within lattice
regularization of QCD [8].

By changing to the center of mass frame
\begin{equation}
\label{neun}
\vec x = \frac{\vec x_1 + \vec x_2}{2}, \quad \vec z = \vec x_1 - \vec x_2
\end{equation}
where $\vec x $ and $\vec z$ are the center of mass and the relative
distances of the $\bar \psi \psi$-system, respectively,
we obtain
\begin{eqnarray}
\label{zehn}
-i S_{QCD}^{(\bar \psi \psi)} &=& -\frac{1}{2} \int d^3 x dt  \int d^3z
V(z) \left[\bar \psi \left( \vec x,t \right) \gamma^{0}
t^A_c
\psi \left( \vec x,t \right) \right] \times \nonumber \\
&& \times \left[ \bar \psi \left( \vec x - \vec z,t \right) \gamma^{0}
t^A_c
\psi \left( \vec x - \vec z,t \right) \right].
\end{eqnarray}
Furthermore, after applying the Fierz transformations
%\newpage
\begin{eqnarray}
\label{elf}
(\gamma^0)_{\alpha \beta} (\gamma^0)_{\gamma \delta} &=&
\frac{1}{4}(1)_{\alpha \delta} (1)_{\gamma \beta} +
\frac{1}{4} (i\gamma^5)_{\alpha \delta} (i\gamma^5)_{\gamma \beta} -
\nonumber \\
&-&\frac{1}{8}
(\gamma_{\mu})_{\alpha \delta} (\gamma^{\mu})_{\gamma \beta} -\frac{1}{8}
(\gamma_{\mu}\gamma^5)_{\alpha \delta} (\gamma^{\mu}\gamma^5)_{\gamma
\beta}
\end{eqnarray}
to the r.h.s. of (\ref{zehn}) and by keeping only the leading terms in the
large-$N$
expansion
\begin{equation}
\label{zwolf}
\sum_{A=1}^{N^2 - 1} \left( t_c^A \right)_{i'j'} \left( t_c^A
\right)_{l'k'} \delta_{ij} \delta_{kl}
\stackrel{N \rightarrow \infty}{\longrightarrow} \sum_{a=0}^{8}
\left( t^a \right)_{il} \left( t^a \right)_{kj} \delta_{i'k'}
\delta_{l'j'}
\end{equation}
eq. (\ref{zehn}) can be written in the form
\begin{eqnarray}
\label{dreizehn}
-i  S_{QCD}^{(\bar \psi \psi)} &=& \frac{1}{8} \int d^4 x   \int d^3 z
V(z) \left[ \bar \psi \left( \vec x,t \right) \Gamma
\psi \left( \vec x- \vec z,t \right) \right] \times \nonumber
\\
&& \times \left[ \bar \psi \left( \vec x - \vec z,t \right)
\Gamma
\psi \left( \vec x,t \right) \right].
\end{eqnarray}
In the previous expression the summation over
$\Gamma=t^a,i\gamma^5 t^a,
\frac{i}{\sqrt{2}}\gamma_{\mu} t^a$ and
$\frac{i}{\sqrt{2}}\gamma_{\mu} \gamma^5 t^a$ has to be performed.

Let us show now that the effective interaction
(\ref{dreizehn}) leads to SBCS and a gap-equation, which determines the
dynamical (constituent) quark mass. We follow the NJL-prescription, i.e. we
accept the one quark-loop approximation, and obtain the effective action
\begin{equation}
\label{vierzehn}
-i \delta S_{QCD}^{eff.mass} = - \frac{1}{8} \int d^4 x   \int d^3 z
\left[ \bar \psi \left( \vec x,t \right)
\psi \left( \vec x -\vec z,t \right) \right] V(z)
\, \mbox{tr}_{C+L} [iS_F(- \vec z)]
\end{equation}
where
\begin{equation}
\label{vierzehna}
\mbox{tr}_{C+L} [iS_F(- \vec z)]= \frac{4mN}{(2\pi)^4 i} \int d^4 k
\frac{e^{-i \vec k \vec z}}{m^2-k_0^2+ {\vec k}^2} = \frac{N}{\pi^2}
\frac{m^2}{z} K_1(mz).
\end{equation}
The McDonald's function $K_1(mz)$ has an exponential behaviour for large
$z$
\begin{equation}
K_1(mz) \stackrel{z \rightarrow \infty}{\longrightarrow} \frac{1}{\sqrt{z}}
e^{-mz}.
\end{equation}
This means that $K_1(mz)$ localizes the integrand of the $\vec
z$-integration in (\ref{vierzehn}) to the region $\mid \vec z \mid \leq
1/m$. Therefore, the wave functions of the dynamical quarks, represented by
the operators $\psi(\vec x,t)$ and $\psi(\vec x-\vec z,t)$, have to be
localized to the regions $\mid \vec x \mid \leq 1/m$ and $\mid \vec x -
\vec z \mid \leq 1/m$ [9], respectively, and we can neglect the $\vec
z$-dependence of $\psi(\vec x-\vec z,t)$, i.e. $\psi(\vec x-\vec z,t)
\simeq \psi(\vec x,t)$. In momentum-space this means that we calculate the
dynamical mass at zero momentum of the dynamical quark ($\vec p=0$) or else
in the rest frame of the dynamical quark. Finally, in the important region
of integration the effective action (\ref{vierzehn}) takes the form
\begin{equation}
\label{achtzehn}
-i \delta S_{QCD}^{eff.mass} = - \int d^4 x \,
\bar \psi \left( x \right)
\psi \left( x \right) \frac{Nm^2}{8\pi^2} \int d^3 z \frac{V(z)}{z}
K_1(mz).
\end{equation}
In eq. (\ref{achtzehn}) the integration over the relative distance $\vec z$
can be performed explicitly
\begin{equation}
\label{neunzehn}
\frac{Nm^2}{8\pi^2} \int d^3 z \frac{V(z)}{z} K_1(mz)= \frac{N \sigma}{\pi}
\frac{1}{m}- \frac{1}{2} N \sqrt{\sigma}.
\end{equation}
Within the NJL-theory the action (\ref{achtzehn}) defines the total value
of the dynamical quark mass. In the framework of this assumption we obtain
the gap-equation
\begin{equation}
\label{zwanzig}
m= \frac{N \sigma}{\pi} \frac{1}{m}- \frac{1}{2} N \sqrt{\sigma},
\end{equation}
which has the following solution
\begin{equation}
\label{einzwanzig}
m= \frac{N \sqrt{\sigma}}{4} \left( \sqrt{1+\frac{16}{\pi N}}-1 \right)
\stackrel{N \rightarrow \infty}{\longrightarrow}
\frac{2\sqrt{\sigma}}{\pi}.
\end{equation}
We want to point out that we obtained the dynamical quark mass at leading
order in the large-$N$ expansion. In other words, we kept only the first
terms in the large-$N$ expansion of the action (\ref{zehn}), and that led
us to the gap-equation (\ref{zwanzig}). With $\sigma=0.27$ GeV$^2$ one gets
for the value of the dynamical quark mass
\begin{equation}
\label{zweizwanzig}
m= \frac{2\sqrt{\sigma}}{\pi} = 0.33 \, \, \mbox{GeV}.
\end{equation}
Our result for the dynamical quark mass agrees with
the predictions of the so-called naive non-relativistic quark model [10]
and with
the value of $m$ within chiral perturbation theory at the quark level based
on the NJL-model calculated in the chiral limit [4].

The emergence of a dynamical quark mass is the result of SBCS. In terms of
$m=2\sqrt{\sigma}/\pi$ we can define the quark condensate, which is the
order parameter for SBCS. In the one quark-loop approximation we obtain
[2-4]
\begin{equation}
\label{dreizwanzig}
\langle 0 \mid \bar \psi \psi \mid 0 \rangle = -\mbox{tr}_{C+L} [iS_F( 0)]
= -4 m I_1(m)
\end{equation}
where $I_1(m)$ is defined in eq. (\ref{vier}). For $\Lambda_{\chi}=0.94$
GeV [4] we get the exact value of the quark condensate [4,11]
\begin{equation}
\langle 0 \mid \bar \psi \psi \mid 0 \rangle = -(0.255 \,\, \mbox{GeV})^3.
\end{equation}
Now we know both the scale of SBCS $\Lambda_{\chi}$ and the dynamical quark
mass $m$. Therefore, starting from the gap-equation (\ref{vier}) we are
able express the constant $G_1$ in terms of the scale of SBCS
$\Lambda_{\chi}$ and the string tension $\sigma$
\begin{equation}
G_1= \frac{1}{\Lambda_{\chi}^2} f \left( \frac{\sigma}{\Lambda_{\chi}^2}
\right)
\end{equation}
where $f ( \sigma/\Lambda_{\chi}^2 )$ is a well known function (see
formulae (\ref{vier}) and (6)).

{}From a different point of view, we can define the constant $G_1$
independently in terms of the string tension $\sigma$ and the confinement
radius $r_c$. Concerning the confinement radius one knows that its value is
of the order of $O(1/m_{\pi})$ where $m_{\pi}=0.14$ GeV is the pion mass.
In our considerations $r_c$ is introduced as an upper limit for the region
of integration over the relative coordinates $\vec z$. In order to express
$G_1$ in terms of the confinement radius we suggest to expand the action
(\ref{dreizehn}) in powers of the relative radius $\vec z$. After this
expansion and the integration over $\vec z$ we obtain
\begin{equation}
-i S_{QCD}^{(\bar \psi \psi)}= \int d^4 x {\cal L}_{NJL} (x) + \int d^4 x
{\cal L}_{(non-NJL)} (x)
\end{equation}
where
\begin{equation}
\label{siebenzw}
{\cal L}_{NJL} (x) = \left( \frac{\sigma \pi}{8} r_c^4 -
\frac{\sqrt{\sigma} \pi}{3} r_c^3 \right) [\bar \psi(x) \Gamma
\psi(x)][\bar \psi(x) \Gamma \psi(x)].
\end{equation}
A comparison between (\ref{siebenzw}) and (\ref{eins}) allows us to relate
$G_1$ to $\sigma$ and $r_c$
\begin{equation}
\label{achtzw}
G_1= \frac{\sigma \pi}{16} r_c^4 - \frac{\sqrt{\sigma} \pi}{6} r_c^3 .
\end{equation}
In principle, using eqs. (\ref{achtzw}) and (25) one can write for the
dependence of the confinement radius $r_c$ on the string tension $\sigma$
and the scale of SBCS $\Lambda_{\chi}$
\begin{equation}
\label{neunzw}
r_c= \frac{1}{\sqrt{\sigma}} \rho \left( \frac{\sigma}{\Lambda_{\chi}^2}
\right) .
\end{equation}
Practically, we insert the phenomenological values of $G_1 \simeq 10$
GeV$^{-2}$ and $\sigma=0.27$ GeV$^{2}$ into (\ref{achtzw}) and estimate the
value of $r_c$. As a result we get $r_c=0.84/m_{\pi}$, which is in
agreement with the restriction $r_c=O(1/m_{\pi})$ explained above.

The Lagrangian ${\cal L}_{(non-NJL)} (x)$ is an infinite series of terms,
which contain derivatives of the quark fields with respect to its
coordinates. These interactions describe the $\bar \psi \psi$-collective
excitations with higher total spins, in comparison to those excitations,
which appear in the NJL-Lagrangian ${\cal L}_{NJL} (x)$. The infinite
number of terms reflects the ability of QCD as the correct theory of the
strong interaction to account for any colorless $\bar \psi \psi$-hadronic
state with quantum numbers of a meson. It is evi\-dent from the
considerations above that ${\cal L}_{(non-NJL)} (x)$ should not contribute
to the gap-equation and, correspondingly, should not influence SBCS.

Thus, in this paper we have shown that the confinement properties of QCD
lead to the same realization of SBCS as in the NJL-model,
and confinement provides the appearance of the local four-quark interaction
of the NJL-kind. The dynamical quark mass and the phenomenological
constants of the NJL-model are expressed in terms of the confinement
parameters $\sigma$ and $r_c$. As a consequence of this result, the ground
states of low-energy QCD and the NJL-model should coincide. This means that
the non-perturbative low-energy QCD vacuum, which is caused by SBCS, can be
fully described by the non-perturbative many-particle vacuum of the
NJL-model, which is evaluated ex\-pli\-cit\-ly in
ref. [2]. This result justifies the assumption of Yoshida concerning the
use of the NJL non-perturbative vacuum wave function as a trial function
for path-integral evaluations in the regime of SBCS within QCD [11]. One
can further conclude that the low-lying $\bar \psi \psi$-colorless
collective excitations within QCD and within the NJL-model should have
similar properties.

The main shortcoming of our considerations is the small value of $G_2$.
Indeed, a comparison between
the Lagrangians (\ref{eins}) and (27) shows that
\begin{equation}
\label{dreissig}
G_2 = \frac{\sigma \pi}{256} r_c^4 - \frac{\sqrt{\sigma}\pi}{12} r_c^3=
\frac{1}{2} G_1,
\end{equation}
i.e. $G_2/G_1=1/2$, instead of the phenomenological value $G_2/G_1=3/2$
[4].
However, this is not crucial for the investigations of this letter, which
has
its main goal in showing the
connection between confinement, expressed by a linear rising potential, and
SBCS. The constant $G_2$ does not
contribute to SBCS and is only responsible for the mass-spectrum of the
vector-meson
nonets. Since we did not intend to
analyse this problem, a detailed discussion of the origin of the value of
$G_2$ goes beyond the scope of this paper. We are planning to touch this
problem in our fourthcoming publications. Shortly, the problem of the value
of
$G_2$ can be settled in different
manners. Within our present approach one could assume
that $G_2$ increases either due to contributions from
${\cal L}_{(non-NJL)}(x)$ or via the terms in the interaction (7), which we
did not take into account, because of their insignificance with respect to
SBCS.
Formally, this means that without loss
of generality one can use the phenomenological value
$G_2 \simeq 15 \mbox{GeV}^{-2}$.
\vskip0.5truecm
Two of the authors
(A.N.I. and N.I.T.) express their gratitude for the warm hospitality at the

Institut f\"ur Kernphysik at Technische Universit\"at Wien, and they thank
for financial support during their stay
in Vienna
when this work was done.

\vglue 0.6cm

\end{document}